# Dynamical Characteristics of Global Stock Markets Based on Time Dependent Tsallis Non-Extensive Statistics and Generalized Hurst Exponents

I.P. Antoniades[a,b], L.P. Karakatsanis[b*], E.G. Pavlos[b]

[a]Division of Technology & Sciences, American College of Thessaloniki, Thessaloniki, Greece.

[b]Complexity Research Team (CRT), Department of Environmental Engineering, Democritus University of Thrace, Xanthi, Greece.

*Corresponding Author: Leonidas P. Karakatsanis, karaka@env.duth.gr

## ABSTRACT

We perform non-linear analysis on stock market indices using time-dependent extended Tsallis statistics. Specifically, we evaluate the $q$-triplet for particular time periods with the purpose of demonstrating the temporal dependence of the extended characteristics of the underlying market dynamics. We apply the analysis on daily close price timeseries of four major global markets (S&P 500, Tokyo-NIKKEI, Frankfurt-DAX, London-LSE). For comparison, we also compute time- dependent Generalized Hurst Exponents (GHE) $H_q$ using the GHE method, thus estimating the temporal evolution of the multiscaling characteristics of the index dynamics. We focus on periods before and after critical market events such as stock market bubbles (2000 dot.com bubble, Japanese 1990 bubble, 2008 US real estate crisis) and find that the temporal trends of $q$-triplet values significantly differ among these periods indicating that in the rising period before a bubble break, the underlying extended statistics of the market dynamics strongly deviates from purely stochastic behavior, whereas, after the breakdown, it gradually converges to the Gaussian-like behavior which is a characteristic of an efficient market. We also conclude that relative temporal variation patterns of the Tsallis q-triplet can be connected to different aspects of market dynamics and reveals useful information about market conditions especially those underlying the development of a stock market bubble. We found specific temporal patterns and trends in the relative variation of the indices in the q-triplet that distinguish periods just before and just after a stock-market bubble break. Differences between endogenous and exogenous stock market crises are also captured by the temporal changes in the Tsallis $q$-triplet. Finally, we introduce two new time-dependent empirical metrics ($Q$-metrics) that are functions of the Tsallis $q$-triplet. We apply them to the above stock market index price timeseries and discuss the significance of their temporal dependence on market dynamics and the possibility of using them, together with the relative temporal changes of the $q$-triplet, as signaling tools for future market events such as the development of a market bubble.

## INTRODUCTION

Complex systems far from equilibrium, present strange dynamics in phase space and can reveal a lot of self-organized phenomena like: strange attractors, chaotic behavior, long range correlations, non-equilibrium state states (NESS), self-organized criticality (SOC), intermittent turbulence, anomalous diffusion, multifractal behavior, non-extensive statistics (Pavlos *et al.*, 2014). Economic systems also present complicated structure and dynamics (Pavlos, 2015). For example, stock market prices are characterized by non-Gaussian distributions (*e.g.* Jondeau, 2007; Iliopoulos, 2015). Many recent works indicated the heavy tails and volatility bursts of real prices. On the other hand, non-extensive statistical mechanics introduced by Tsallis (Tsallis, 2009) based on a generalization of Boltzmann – Gibbs entropy describe far from equilibrium non-linear complex dynamics which reveal self-organization, multifractal strange attractors and long-range correlations. The non-extensive statistics of Tsallis is based on the *q-entropy* function, the maximization of which can reproduce geometrical, dynamical and statistical characteristics of non-equilibrium complex dynamics.

[1]

In this direction, Stosic *et al.* (Stosic, 2019) studied the non-extensivity of price volatilities for 34 major stock market indices between 2010 and 2019 by the estimation of *q*-triplet of Tsallis statistics. The distances between *q*-triplets indicated that some stock markets might share similar non-extensive dynamics, while others are widely different. The study concentrated in comparing non-extensive statistics between major stock markets for a particular time-period only. Trindade *et al.* (Trindade, 2020) developed a strategic of optimal portfolio based on information theory and Tsallis statistics, showing advantages of the optimal portfolio over an arbitrary choice of causal portfolios. The wealth after *n* days with the optimal portfolio is given by a *q*-exponential function. Moreover, Wang and Shang (Wang, 2018) proposed a multiscale cross distribution entropy (MCDE) method based on the Tsallis entropy to analyze financial stock markets. The results showed that the embedding dimension *m* has little influence on MCDE and the stability of financial timeseries could be affected by the order *q*. In addition, Batra, L and Taneja, HC (Batra, 2020) used the information theoretic approach to analyze the stock indices and sectors which are highly volatile. They have used entropy measures like: Shannon, Tsallis and Renyi entropy and the Approximate & Sample entropy and they concluded that information theoretic measures can be useful to characterize the volatility in financial markets. Furthermore, Zhang *et al.* (Zhang, 2020) proposed a method based on multi-scale permuted distribution Cumulative Tsallis entropy (MPDCTE) to analyze the complexity and dissimilarity between data. They applied the method in simulation data and real financial markets from nine selected stocks. The MPDCTE method clearly divided the stocks by analyzing the data, which is consistent with the phenomenon of the financial market.

The above studies looked into, compared and classified financial markets by studying the non-extensive characteristics of financial timeseries coming from a single time period. In the present study, however, we attempt to further this investigation by focusing on the *temporal element* of complex behavior of the stock market dynamics by looking into changes in the multifractal and non-Gaussian character as well as, the non-extensivity of the statistics of financial timeseries evaluated in different time periods. These periods were chosen so that they are relevant to the market, *i.e.* they correspond to eras marked by well-known market events, such as the rising and dropping period of stock market bubbles and financial crises. By this, we aim to link relative changes in the non-extensive statistics in each chosen time period to known financial features of these periods and thus look into the possibility that Tsallis statistics can be used as a signaling (warning) tool for future market events, such as the development of a stock market bubble. Concerning the methods we used, for each chosen time period, we estimated the Tsallis *q*-triplet ($q_{sen}$, $q_{rel}$, $q_{stat}$), corresponding to three distinct physical processes: *q*-entropy production, *q*-relaxation and the scale invariant meta-equilibrium stationary states in respect. In addition, for each of the chosen time periods, we applied the Generalized Hurst Exponent (GHE) methodology in order to study the temporal changes of GHE-based multiscaling metrics. This allowed us to compare and contrast the insight gained on market dynamics by the Tsallis *q*-triplet and GHE approaches combined. Finally, we introduced two new empirical metrics (that we call *Q* metrics) that depend on the Tsallis *q*-triplet that can be used as measures of the temporal evolution of *total* non-extensivity of the statistics. We applied all the above methods to four major stock markets (S&P 500, Tokyo-NIKKEI, Frankfurt-DAX and London-LSE), specifically to the historical daily close price timeseries from these markets.

## THEORETICAL FRAMEWORK

### Non-extensive statistical mechanics

The non-extensive statistical theory is based mathematically on the nonlinear equation:

$$\frac{dy}{dx} = y^q, y(0) = 1, q \in \mathrm{R}, \tag{1}$$

which has as solution the $q$-exponential function defined by: $e^x = [1 + (1-q)x]^{\frac{1}{1-q}}$. In order to characterize the non-Gaussian character of the dynamics, one estimates the Tsallis $q$-triplet, based on Tsallis non-extensive statistical mechanics. Non-extensive statistical mechanics includes the $q$-analog (extension) of the classical Central Limit Theorem (CLT) and $\alpha$-stable distributions corresponding to dynamical statistics of globally correlated systems. The $q$-extension of CLT leads to the definition of statistical $q$-parameters of which the most significant is the $q$-triplet ($q_{stat}, q_{sen}, q_{rel}$), where the abbreviations *sen*, *rel*, and *stat*, stand for *sensitivity* (to the initial conditions), *relaxation* and *stationary* (state) in non-extensive statistics respectively (Tsallis, 2004, 2011; Umarov, 2008). These quantities characterize three physical processes: a) $q$-entropy production ($q_{sen}$), b) relaxation process ($q_{rel}$) of an induced disturbance of the system to steady state, and c) fluctuations ($q_{stat}$) of the system at equilibrium (steady) state. The $q$-triplet values characterize the attractor set of the dynamics in the phase space of the dynamics and they can change when the dynamics of the system is attracted to another attractor set of the phase space. Equation (1) for $q = 1$ corresponds to the case of equilibrium Gaussian (BG) world (Tsallis, 2009). In this case, the $q$-triplet of Tsallis simplified to $q_{stat} = 1, q_{sen} = 1, q_{rel} = 1$.

### $q$-triplet of Tsallis theory

In this subsection we present a short summary of the theoretical background for each member of the Tsallis $q$-triplet.

*a) $q_{stat}$ index*

A long-range-correlated meta-equilibrium non-extensive process can be described by the nonlinear differential equation (Tsallis, 2004):

$$\frac{d(p_i Z_{q_{stat}})}{dE_i} = -\beta_{q_{stat}} (p_i Z_{q_{stat}})^{q_{stat}} \tag{2}$$

The solution of (2) corresponds to the probability distribution:

$$p_i = e_{q_{stat}}^{-\beta_{q_{stat}} E_i} / Z_{q_{stat}} \tag{3}$$

where $\beta_{q_{stat}} = \frac{1}{KT_{stat}}$ and $Z_{q_{stat}} = \sum_j e_{q_{stat}}^{-\beta_{q_{stat}} E_j}$ is the partition function. Then, the probability distribution is given by:

$$p_i \propto \left[1 - (1-q)\beta_{q_{stat}} E_i\right]^{\frac{1}{q_{stat}-1}} \tag{4}$$

for discrete energy states $\{E_i\}$ and by

$$p(x) \propto \left[1 - (1-q)\beta_{q_{stat}} x^2\right]^{\frac{1}{q_{stat}-1}} \tag{5}$$

for continuous $x$ states $\{X\}$, where the values of the magnitude $X$ correspond to the state points of the phase space. Distribution functions (4) and (5) correspond to the attracting stationary solution of the extended (anomalous) diffusion equation related to the nonlinear dynamics of the system. The stationary solutions $p(x)$ describe the probabilistic character of the dynamics on the attractor set of the phase space. The non-equilibrium dynamics can evolve on distinct attractor sets, depending upon the control parameters, while the $q_{stat}$ exponent can change as the attractor set of the dynamics changes. In the present work, for the estimation of Tsallis $q$-Gaussian distributions we use the method described in Ferry *et al.* (Ferry, 2010).

*b) $q_{sen}$ index*

Entropy production is related to the general profile of the attractor set of the dynamics. The profile of the attractor can be described by its multifractality as well as by its sensitivity to initial conditions. The sensitivity to initial conditions can be expressed as:

$$\frac{d\xi}{dt} = \lambda_1 \xi + (\lambda_q - \lambda_1)\xi^q \tag{6}$$

where $\xi$ is the trajectory deviation in the phase space: $\xi \equiv \lim_{\Delta x \to 0} [\Delta x(t)/\Delta x(0)]$ and $\Delta x(t)$ is the distance between neighboring trajectories (Tsallis, 2004). The solution of equation (6) is given by:

$$\xi = \left[1 - \frac{\lambda_{q_{sen}}}{\lambda_1} + \frac{\lambda_{q_{sen}}}{\lambda_1} e^{(1-q_{sen})\lambda_1 t}\right]^{\frac{1}{1-q_{sen}}} \tag{7}$$

The $q_{sen}$ exponent is related to the multifractal profile of the attractor set according to

$$\frac{1}{1-q_{sen}} = \frac{1}{\alpha_{min}} - \frac{1}{\alpha_{max}} \tag{8}$$

where $\alpha_{min}$, $\alpha_{max}$ corresponds to the zero points of the multifractal exponent spectrum $f(\alpha)$, i.e. $f(\alpha_{min}) = f(\alpha_{max}) = 0$. For the estimation of the multifractal spectrum we use the method described in Pavlos *et al.* (Pavlos, 2014).

*c) $q_{rel}$ index*

Thermodynamic fluctuation – dissipation theory is based on the Einstein original diffusion theory (Brownian motion theory). Diffusion is a physical mechanism for extremization of entropy. If $\Delta S$ denotes the deviation of entropy from its equilibrium value $S_0$, then the probability $P$ of a proposed fluctuation is given by:

$$P \approx \exp(\Delta S/k) \tag{9}$$

The Einstein – Smoluchowski theory of Brownian motion was extended to the general FP diffusion theory of non-equilibrium processes. The potential of FP equation may include many meta-equilibrium stationary states near or far away from thermodynamical equilibrium. Macroscopically, relaxation to the equilibrium stationary state of some dynamical observable $O(t)$ related to system evolution in phase space can be described by the general form:

$$\frac{d\Omega}{dt} \cong -\frac{1}{\tau}\Omega, \tag{10}$$

where $\Omega(t) = [O(t) - O(\infty)]/[O(0) - O(\infty)]$ describes the relaxation of the macroscopic observable $O(t)$ towards its stationary state value, $\tau$ being the relaxation time (Tsallis, 2004). The non-extensive generalization of fluctuation – dissipation theory is related to the general correlated anomalous diffusion processes (Tsallis, 2009). The equilibrium relaxation process is transformed to the meta-equilibrium non-extensive relaxation process according to:

$$\frac{d\Omega}{dt} \cong -\frac{1}{\tau_{q_{rel}}}\Omega^{q_{rel}}. \tag{11}$$

The solution of (11) is given by:

$$\Omega(t) \cong e_{q_{rel}}^{-t/\tau_{q_{rel}}} \tag{12}$$

The autocorrelation function $C(t)$ or the mutual information $I(t)$ can be used as candidate observables $\Omega(t)$ for the estimation of $q_{rel}$ (Pavlos EG, 2019). However, in contrast to the linear profile of the correlation function, the mutual information includes the nonlinearity of the underlying dynamics and is proposed as a more faithful index of the relaxation process and thus for the estimation of the Tsallis exponent $q_{rel}$.

## Generalized Hurst Exponent ($H_q$)

The Hurst exponent (Hurst, 1951; 1965) is a well-known tool used to study the scaling behavior of timeseries coming from any dynamical process. To compute the scaling exponents, it is necessary to study the *q*-order moments of the absolute value of the increments of the timeseries (Di Matteo, 2007). In particular, a timeseries $X_t$ with stationary increments is analyzed through the structure function $S_q$:

$$S_q = \langle |X_{n+\tau} - X_n|^q \rangle_n \sim \tau^{qH_q}$$

(13)

where $q>0$ is a real parameter (not to be confused with the Tsallis $q$-indices), $\tau$ is the lag time and $\langle ... \rangle_n$ is the average over all time steps $n$ of the series. $H_q$ is called the *Generalized Hurst Exponent* (GHE). The function $qH_q$ is concave (Mandelbrot 1963; 1997) in $q$ and codifies the scaling exponents of the process. $H$ is the historical (basic) definition of the Hurst exponent, which corresponds to the $H_q$ spectrum value for $q = 1$. If $H_1 = 0.5$, the evolution of the system in state-space is equivalent to a random walk, *i.e.* the underlying process is purely stochastic (diffusive). For a single variable timeseries, this is equivalent to saying that at any given time, the value of the series is equally likely to go up as it is to go down. For $H_1 > 0.5$, the system evolves faster than stochastic diffusion (super-diffusive process), which implies that -for a single-variable series- if a change occurs in one direction (up or down), it is more likely that the next change will be in the same direction rather than in the opposite. In such a case, the underlying process is characterized as a *persistent* process. Finally, for $H_1 < 0.5$, the system evolves slower than stochastic diffusion (sub-diffusive process). For a single-variable series, this implies that, if a change occurs in one direction (up or down), it is more likely that the next change will be in the opposite direction. In the latter case, the process is characterized as an *anti-persistent* process. For the calculation of GHE's of higher and more positive values of $q$, the largest differences in a series are weighted more than smaller differences in equation (13) and therefore large-$q$ GHE's emphasize the *tails* of the distribution of differences. Conversely, lower (less positive) values of $q$ weigh small differences more than large ones. Computing a broad spectrum of GHE's, for several spread-out values of $q$, provides a more detailed 'signature' of the underlying dynamics of the system compared to considering only the original Hurst exponent. For computing the $H_q$ values, we use the GHE method with a maximum value of time lag $\tau$ equal to 19 trading days, as prescribed in (Di Matteo, 2003; 2005).

For *small q* (0<q<~2), a multiscaling proxy can be obtained from GHE's by fitting the measured scaling exponents with a second degree polynomial of the form[1]

$$qH_q \cong Aq + Bq^2$$

(14)

or equivalently (Brandi, 2020):

$$H_q \cong A + Bq,$$

(15)

where $A$ and $B$ are two constants. In this setting, $B$, which can simply be determined by a least squares linear fit to $H_q$ vs. $q$ data, represents the curvature of $qH_q$. We note that the linear assumption for the form of $H_q$ is very well justified in the present study as we used a range for $q$-values in the interval [0.1,1]. If $B = 0$, then $H_q$ does not depend on $q$, *i.e.* $H_q = H$ for all $q$, hence the process is *uniscaling*, while if $B \neq 0$, the process is multiscaling (Di Matteo, 2007; Buonocore, 2016; 2020; Brandi, 2020). Thus, $B$, to which, from here on, we will refer as *B-proxy*, or *GHE spectrum depth* is a measure of the degree of multiscaling of a timeseries.

Except for the *B*-proxy, another measure of the degree of multiscaling in each chosen time period that we used is the *width* $W_{q,q'}$ of the GHE spectrum defined as

$$W_{q,q'} = H_q - H_{q'},\tag{16}$$

where $q < q'$. For $W_{q,q'}$ close to zero, the underlying series is uniscaling, whereas the larger $W_{q,q'}$ the more multiscaling the timeseries. Parameters $q, q'$, should be chosen as an extreme pair of $q$ values (*e.g.* $q \leq 1, q' > 2$) in order to obtain a broad enough weighting of the full range of the distribution of differences. However, using high values for $q'$ can bias the results, if the data is characterized by distributions with fat tails. In particular, for $q > \alpha$, where $\alpha$ is the tail exponent of the difference distribution of the data, the $q$-moments are not well defined. This introduces a bias on the expected value which in turn, produces a bias in the GHE estimation. Since financial timeseries are generally fat-tailed, the choice of $q$ is relevant for the two measures of multiscaling, the width $W$ and the depth $B$. One possible choice, that was suggested recently by Antoniades *et al.* in (Antoniades, 2020), is to pick a narrow range of $q$-$q'$ values for $B$ and a wide range for $W$. This way, the value of $B$ is not so much biased by the extreme edges of the difference distributions, but is rather affected by the small and medium differences. $W$, on the other hand, is deliberately allowed to be affected by the extreme tail data, which in financial timeseries are potentially important. Then, $B$ and $W$ can both be used in conjunction as measures of multiscaling, each of them weighing a different range in the difference distributions and complementing each other in conveying useful information about the scaling of the timeseries. Hence, in the present work, following the practice of (Antoniades, 2020), we used $q=0.1$ and $q'=4$ for $W$ and $q=0.1$ and $q'=1$ for $B$.

---

[1]Technical details of the choice of this functional form can be found in (Buonocore, 2016; 2020).

[6]

Apart of multiscaling, a measure of the degree of multifractality of a timeseries that has extensively been used in literature is the width of the multifractal spectrum $f(\alpha)$ *vs.* $\alpha$, defined as the difference between the maximum and minimum values of $\alpha$:

$$\Delta\alpha = \alpha_{max} - \alpha_{min},\tag{17}$$

One can easily check from (8) that $\Delta\alpha$ is related to the index $q_{sen}$ via:

$$\frac{1}{1-q_{sen}} = \frac{\Delta\alpha}{\alpha_{max} \cdot \alpha_{min}}\tag{18}$$

*i.e.* $\Delta\alpha$ is inversely proportional to $1 - q_{sen}$.

**$q$-norm and $q$-distance**

One can define the norm $\|\vec{q}\|$ of the $q$-vector $\vec{q} = (q_{stat}, q_{sen}, q_{rel})$ by

$$\|\vec{q}\| = \sqrt{q_{stat}^2 + q_{sen}^2 + q_{rel}^2}\tag{19}$$

and use it as a characteristic measure of the 'strength' of the 'non-extensivity' of the statistics of a complex system; the larger the norm, the more non-extensive the statistics, *i.e.* the more the statistics of the underlying system departs from randomness. Moreover, one can define the Euclidean distance $d_{q,q'}$ between the $q$-vectors $\vec{q}$ and $\vec{q}'$ of two timeseries and use it as a measure of the total difference between the non-extensive statistics of the system:

$$d_{\vec{q},\vec{q}'} = \sqrt{(q_{stat} - q'_{stat})^2 + (q_{sen} - q'_{sen})^2 + (q_{rel} - q'_{rel})^2} \tag{20}$$

The quantity defined by equation (20) is known as the *q-distance* between two *q*-triplets and has been used before in financial systems (for example in Stosic, 2019) to describe differences between the extensive statistics of two distinct timeseries.

### Normalized *W* and *B*

As mentioned in the introduction, both W and B as determined from the GHE spectrum are measures of the degree of multiscaling in the timeseries, each one focusing on a different part of the difference distributions and conveying different information about multiscaling. In order for their values to be comparable, for each time period, we choose a particular normalization that will bring them to the same scale. One type of normalization would be to divide them by using the average and standard deviation of the set of *W* and *B* values for all time periods. However, we applied a different type of normalization, as introduced in (Antoniades, 2020), and shown below:

$$W'(t_i) = \frac{W(t_i) - \langle W^{surr}(t_i)\rangle}{\sigma(W^{surr}(t_i))} \tag{21}$$

$$B'(t_i) = \frac{B(t_i) - \langle B^{surr}(t_i)\rangle}{\sigma(B^{surr}(t_i))} \tag{22}$$

where $W^{surr}(t_i)$ or $B^{surr}(t_i)$ is the respective *W* or *B* of the *i*th-period of a *random surrogate* timeseries having exactly the same length as the *i*th-period of the original series, but being generated with % changes drawn from a normal distribution. $\sigma(W^{surr}(t_i))$ or $\sigma(B^{surr}(t_i))$ denotes the *standard deviation* of the set of $W^{surr}(t_i)$ or $B^{surr}(t_i)$ over all time periods $i = 1,2,...,n$ of the random surrogate series. The rationale of above type of normalization is that its value directly relates *W* and *B* to the temporal variability of the same quantities of a randomly generated timeseries which ideally yields $W = B = 0$, but for any short-length time period this is not the case due to *finite-size effects*. Thus, $\sigma(W^{surr}(t_i))$ and $\sigma(B^{surr}(t_i))$ essentially measure the 'noise-level' in the GHE calculations that is attributed to finite-size effects. Therefore, the values of $W'$ and $B'$ express the number of standard deviations above the noise level that the values of *W* and *B* represent for the real timeseries and thus provide a measure of the statistical significance of the observed multiscaling. A second advantage of normalizing relative to a random surrogate series is that it provides a single reference point and thus allows comparisons of multiscaling strength among different timeseries. The latter is not possible if, for example, the temporal variability of $W'$ and $B'$ in the real timeseries is used for normalization.

### *Q*-metrics: Two new empirical metrics based on Tsallis *q*-triplet:

As mentioned in previous paragraphs, in the case of a random timeseries $q_{stat} = q_{sen} = q_{rel} = 1$. Therefore, the more each member of the *q*-triplet of a dynamical system deviates from unity, the more the statistics of the dynamical system deviates from randomness. In this sense, the Euclidean distance of any $q$-triplet from the point (1,1,1) can be used as a measure of deviation from randomness:

$$D_{\vec{q}} = d_{q,1} = \sqrt{(q_{stat} - 1)^2 + (q_{sen} - 1)^2 + (q_{rel} - 1)^2} \tag{23}$$

[8]

However, in the present work our goal is to compare $q$-triplets evaluated for different time periods of the same timeseries. Each time period $i$ starts at time $t_i - \Delta t_i$ and ends at time $t_i$ ($i = 1,2,...,n$), where $\Delta t_i$ is the duration of the period. As each member of the triplet potentially varies on different scales (usually $q_{stat}$ varies in the range 1-2, $q_{sen}$ in the range -3 to 1 but $q_{rel}$ may obtain much higher values such as 5-10 or higher), temporal differences in the values of a metric involving all members of the triplet, like the one in equation (23), may not be equally affected by each member of the triplet, since its values maybe overwhelmed by the member of the $q$-triplet with the largest absolute values. Therefore, to amend this issue, we use a normalization trick that brings all members of the $q$-triplet to the same scale, while at the same time the absolute value of the metric is a measure of the total deviation of the statistics from randomness. The following empirical metric is thus obtained:

$$Q(t_i) = \frac{1}{\sqrt{3}} \langle D_{\vec{q_i}} \rangle \sqrt{\left(\frac{q_{stat}(t_i)}{\langle q_{stat}(t_i) \rangle}\right)^2 + \left(\frac{q_{sen}(t_i)}{\langle q_{sen}(t_i) \rangle}\right)^2 + \left(\frac{q_{rel}(t_i)}{\langle q_{rel}(t_i) \rangle}\right)^2}, \tag{24}$$

where $\vec{q_i} = (q_{stat}(t_i), q_{sen}(t_i), q_{rel}(t_i))$ is the $q$-triplet vector for time period $[t_i - \Delta t_i, t_i]$ ($i = 1,2,...,n$), $D_{\vec{q_i}}$ is given by (23) and $\langle X \rangle$ denotes average of the contained quantity over all time periods $i$:

$$\langle X \rangle = \frac{\sum_i X_i}{n} \tag{25}$$

The square root expression in (24) is essentially the norm of the $q$-triplet vector for each time period $i$, whose coordinates are normalized by the average value of the respective member of the triplet over all time periods the timeseries is divided into. It can be seen that, if $\vec{q}(t_i) = (\langle q_{stat}(t_i) \rangle, \langle q_{sen}(t_i) \rangle, \langle q_{rel}(t_i) \rangle)$, then $Q(t_i) \equiv \langle D_{\vec{q_i}} \rangle$. In this sense, $Q(t_i)$ is a rescaled version of $D_{\vec{q}(t_i)}$ (as given by equation (23)), in which each member of the triplet $\vec{q}(t_i)$ is normalized by its average value.

An alternative metric to (24), with a different physical interpretation can also be introduced based on the following rationale: Since $\Delta\alpha \propto 1/(1 - q_{sen})$, as seen by (18), and since $\Delta\alpha$ measures the degree of multifractality of a timeseries, it follows that the more multifractal the timeseries, the smaller $1 - q_{sen}$ and vice-versa. This implies that the value of the $Q$-metric defined by (24) is larger for unifractal timeseries and smaller the more multifractal a timeseries. However, the more random a timeseries the more unifractal its scaling, i.e. $\Delta\alpha \to 0$ or equivalently, $1 - q_{sen} \to \infty$. This can be counter-intuitive, as one expects a more multifractal timeseries to be considered as being farther away from 'randomness'. In fact several studies of the temporal evolution of multiscaling in stock-market price timeseries (e.g. Antoniades, 2020; Morales, 2012; Drożdż, 2015, 2018) have shown that in periods of a market crisis, stock index price timeseries become more multiscaling, whereas in efficient market periods they become more uniscaling. (For a further review of studies of multiscaling in financial timeseries the interested reader may also refer to (Jiang, 2019; Di Matteo, 2003, 2005, 2007; Mandelbrot, 1963, 1997; Buonocore, 2016, 2020). Therefore, if we replace

[9]

$q_{sen}$ in (23) by $1/(1 - q_{sen})$, we define a type of '$q$-distance' for any $q$-triplet vector $\vec{q}$, which we call $D_{\vec{q}}^{inv}$:

$$D_{\vec{q}}^{inv} = \sqrt{(q_{stat} - 1)^2 + \left(\frac{1}{q_{sen}-1}\right)^2 + (q_{rel} - 1)^2} \tag{26}$$

We can then modify (24) accordingly in order to obtain the equivalent of the $Q$-metric that we will call $Q^{inv}$-metric:

$$Q^{inv}(t_i) = \frac{1}{\sqrt{3}} \langle D^{inv} \rangle \sqrt{\left(\frac{q_{stat}(t_i)}{\langle q_{stat}(t_i)\rangle}\right)^2 + \left[\frac{(1-q_{sen}(t_i))^{-1}}{\langle(1-q_{sen}(t_i))^{-1}\rangle}\right]^2 + \left(\frac{q_{rel}(t_i)}{\langle q_{rel}(t_i)\rangle}\right)^2}, \tag{27}$$

## RESULTS

### Data description

For our analysis, we have used 4 stock market indices: New York stock exchange index (S&P 500), Tokyo stock exchange index (NIKKEI), Frankfurt stock exchange general index (DAX) and the London-LSE. Table 1 shows the time period in which the data is analyzed and the total number of trading days in each series.

Table 1: Time periods and the number of trading days analyzed for each stock market.

| Market | Time period | Trading days |
|---|---|---|
| S&P 500 | 1955-2020 | 16,377 |
| NIKKEI | 1965-2020 | 13,068 |
| DAX | 1987-2020 | 7,992 |
| LSE | 2001-2020 | 4,699 |

Next, we have divided each timeseries into a number of uneven segments whose limits roughly correspond to the beginning and the end of particular market periods, including critical periods, such as the beginning of a steep price rise or drop. The precise break-down into time-periods for each index is shown later on in Tables 2-5.

### Results of Tsallis $q$-triplet, $W$, $B$ and $Q$-metric calculations

In this subsection we present the results of the Tsallis statistics and GHE calculations applied to the S&P 500, NIKKEI, DAX and LSE indices. In Tables 2-5 we show the limits of the time periods for which we computed the Tsallis $q$-triplet for each index. The beginning of the next time period is always the end of the previous period plus one trading day, therefore the entire historical timeline of the index (for which we had available data) is covered and there are no overlapping periods. Some of the periods were chosen so that they roughly correspond to the characteristic periods before and after a particular market crisis, such as the 2000 dot.com bubble, the Japanese 1990-1991 bubble and the 2008 USA real estate crisis, in order to capture the variation of the $q$-triplet as well as its corresponding $Q$-metric values and the values of $W$ and $B$ before and after the crises.

In figures 1-4, for each of the above indices, we show $q_{stat}(t_i)$ together with the index daily close prices in plots (a), $1 - q_{sen}(t_i)$ in plots (b), $q_{rel}(t_i)$ in plots (c), the respective values $Q(t_i)$ and $Q^{inv}(t_i)$ of the $Q$-metrics in plots (d) and finally the normalized width $W'$ and normalized depth $B'$ of the GHE spectrum in plots (e). $t_i$ is the *right* end of each period $i$; the values of all computed quantities are shown by vertical bar graphs, where the bars are positioned *at the end* $(t_i)$ of each period $i$. The error bars shown in figures (a), (b), (c) and (e) represent the uncertainties in the respective quantities as estimated by the fitting procedures, whereas the error bars for the $Q$-metrics shown in figures (d) are calculated by error propagation from equations (24) and (27). The $P$-values for the linear regressions yielding the $B$-proxy are all usually over 0.98 and at worst 0.95, indicating that equation (15) is a very good approximation. In all figures we include linear least-squares fits, for each member of the $q$-triplet, performed on data both before a particular crisis, denoted by striped black lines. These are included in order to show temporal trends of the q-triplet before a crisis. The figures also include the *local* trends before and after the 2000(S&P 500, DAX), 1990 (NIKKEI) and 2008 (S&P 500, DAX, NIKKEI, LSE) crises that are shown by solid red arrows. The significance of the local and long-term temporal trends is discussed later in this section and in the discussion section.

*S&P 500 Stock Market*

Table 2 shows the periods for which the $q$-triplet, $W$, $B$ and the two $Q$-metrics were computed for the S&P 500 index. In the last column of the table an explanation of the market event corresponding to the particular date is given. Shaded cells correspond to dates with no particular known event, which were chosen so that neighboring time periods contain approximately the same number of trading days.

Table 2: Description of the break-down of S&P 500 close price timeseries in time periods used for the Tsallis statistics, GHE's and $Q$-metrics calculations.

| Period No | Duration (Tr. Days) | Start/End | Dates | Related market event |
|---|---|---|---|---|
| 1 | 2062 | Start | 21/1/1955 | Start of recorded data |
| | | End | 29/3/1963 | |
| 2 | 2030 | Start | 30/3/1963 | |
| | | End | 28/5/1971 | (roughly) The end of Hurst exponent rising trend[2] |
| 3 | 2063 | Start | 1/6/1971 | " |
| | | End | 31/7/1979 | (roughly) 2nd oil crisis of the 70's |
| 4 | 2066 | Start | 1/8/1979 | " |
| | | End | 1/10/1987 | Three weeks before the 'Black Monday' crash event |
| 5 | 1822 | Start | 2/10/1987 | " |
| | | end | 14/12/1994 | (roughly) Beginning of the year 2000 dot.com bubble rising period. |
| 6 | 1132 | Start | 15/12/1994 | " |

---

[2] The rising trend in the time-dependent Hurst exponent ($H_1$) of S&P 500 has been reported in (Alvarez 2008) using a sliding time window for the period 1950-1972. The rising trend was followed by a dropping trend extended in the period 1972-2008. *See* fig. 6 and 7 in the said study. This trend was also observed in the Hurst calculations performed in this study.

[11]

| | | End | 10/6/1999 | (roughly) End of the year 2000 dot.com bubble rising period. |
|---|---|---|---|---|
| 7 | 1068 | Start | 11/6/1999 | (roughly) Beginning of the year 2000 dot.com bubble crash. |
| | | End | 10/9/2003 | (roughly) End of the year 2000 dot.com bubble crash |
| 8 | 1043 | Start | 11/9/2003 | (roughly) Beginning of mid-millennia's blooming period |
| | | End | 31/10/2007 | (roughly) End of mid-millennia's blooming period |
| 9 | 337 | Start | 1/11/2007 | (roughly) Beginning of 2008 US real estate related market crash |
| | | End | 5/3/2009 | (roughly) End of 2008 US real estate related market crash |
| 10 | 1628 | Start | 6/3/2009 | " |
| | | End | 21/8/2015 | |
| 11 | 1127 | Start | 22/8/2015 | |
| | | End | 13/2/2020 | End of recorded data |

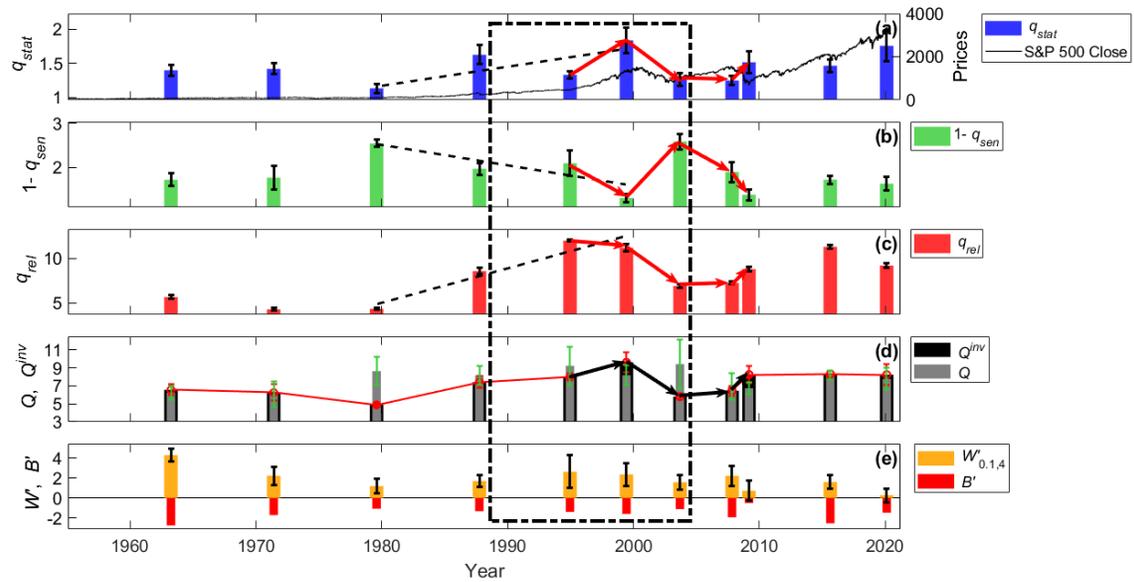

Figure 1 shows the results for the S&P 500 index, as discussed above.

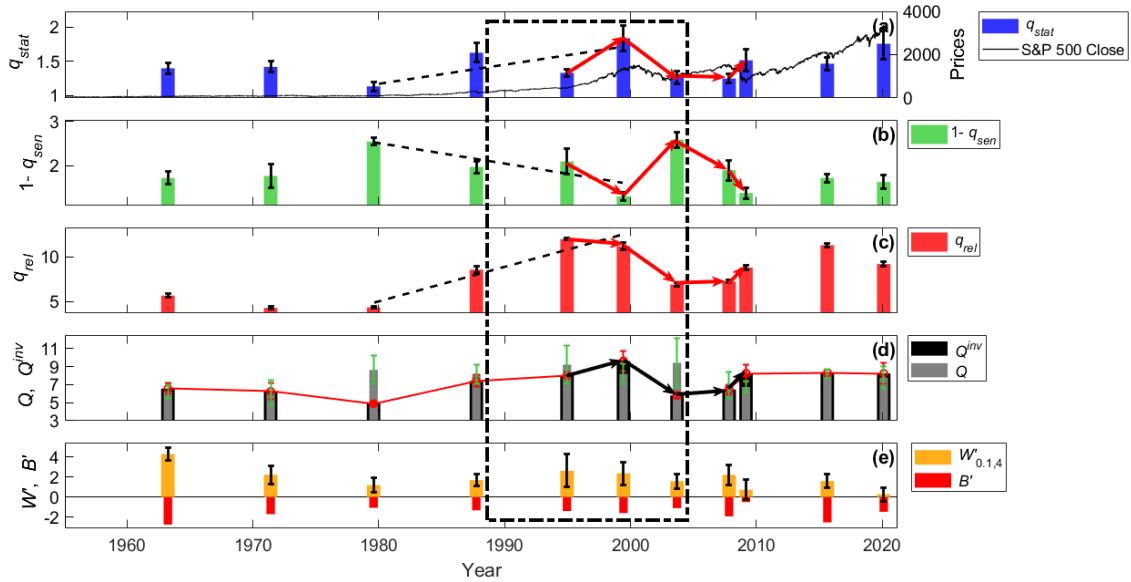

Figure 1: Time-dependent Tsallis $q$-triplet, $Q$-metrics, normalized GHE width $W'$ and normalized $B$-proxy $B'$ for the S&P 500 stock market daily closing price timeseries for the time-periods shown in Table 2. a) $q_{stat}$, b) $q_{sen}$ c) $q_{rel}$, d) $Q$ and $Q^{inv}$, (e) $W'_{0.1,4}$, $B'_{0.1,1}$. The striped box encloses two periods before the 2000 bubble break and the period after the break. The red arrows in (a), (b), (c) show the local trends in the $q$-triplet before and after the bubble break for the 2000 and the 2008 crises and the black arrows in (d) the respective trends in $Q^{inv}$. The solid red line in (d) is a line plot of metric $Q^{inv}$ and serves merely as a guide to the eye showing all the local trends in $Q^{inv}$. The striped black lines show linear least squares fits to the data before the year 2000 crisis for each index of the $q$-triplet.

We see that before the break of the 2000 dot.com bubble, during the market rising period, $q_{stat}$ gradually rises to a value of $1.84 \pm 0.44$ showing a clear departure from Gaussian statistics and it rapidly drops down to $1.26 \pm 0.10$ in the period after the bubble break. As $q_{stat}$ is related to the deviation of the price % change distribution from a Gaussian and is particularly affected by the 'fat tails', its increased value in the rising period of the year 2000 dot.com bubble is a consequence of the frequent large change events that occurred during that period. One such big tail event was the 'Black Monday' crash, in which the S&P 500 index lost more than 20% in a single day, followed by two large rises (+5.3% and +9.1%) in the next couple of days and a prolonged period of instability where large drops were followed by large rises within a few days. In the same period, there was also the so called 1997 'Asian crisis' and the 1998 'Russian crisis', both of which yielded a few very large tail events. The characteristic of all these events is that big drops were followed by big rises. However, except for these isolated extreme events, in these periods there is an abundance of medium-sized daily % price change events that show this anti-persistent characteristic: big drops, most likely followed by big rises. This is a type of market behavior that is a consequence of a 'nervousness' among traders, characteristic in periods preceding a stock-market bubble, as discussed recently in (Antoniades 2020) (*see* the discussion section therein). In the same period, the increased $q_{stat}$ value is accompanied by an increased $W$ value, signifying strongly multiscaling dynamics, an observation that is in accordance to the small $q_{sen}$ value; the smaller $q_{sen}$ the wider the multifractal spectrum $\Delta\alpha$ and the larger the GHE spectrum width $W$. In the same period, a rising trend in $q_{rel}$ is observed leading to a notably large value ($11.2 \pm 0.45$) in the period preceding the 2000 dot.com crash. This is an indication that the underlying dynamics is characterized by larger relaxation times, if disturbed, which means that an orbit in phase space lies on a more low-dimensional attractor, which in turn indicates less random and more coordinated market behavior during the market rise before the year 2000; this dynamical behavior is alluding to a market trading phenomenon known as 'herding' in financial jargon. Our result agrees with other studies that have noted herding behavior in critical periods in some markets including S&P 500. In conclusion, the period before the 2000 dot.com bubble crash is characterized by highly non-extensive statistics as indicated by all three Tsallis indices, with a particular temporal trend: *rising* $q_{stat}$, *dropping* $1 - q_{sen}$ and a *rising* $q_{rel}$. This fact is also very clearly demonstrated by the *rising* trend in the $Q^{inv}$-metric, since this metric is a *total* measure of the non-extensivity of the dynamics summing the contributions from each member of the $q$-triplet. In the next period, during the dot.com crash, we observe a notable reduction in the $q_{stat}$ index, which obtains its lowest value of $1.26 \pm 0.10$, denoting a more Gaussian price change distribution during the prolonged period of the crash. The same notable reduction is observed in $q_{rel}$, which indicates that there is a clear transition of the dynamics to a higher dimensional space, as the dynamics becomes more stochastic. Also, the significant rise of $1 - q_{sen}$ and equivalently the significant drop of *W* and *B*, indicate that we move to an almost uniscaling market. As a consequence of all the above, there is a significant drop in the $Q^{inv}$-metric. Notice that the $Q$-metric value, on the other hand, remains high, since the drop in $q_{stat}$ and $q_{rel}$ is overwhelmed by the large increase in $1 - q_{sen}$.

[13]

It is interesting to notice the difference in the temporal trends of the $q$-triplet before and after the externally caused 2008 US real estate crisis compared to the dot.com bubble in 2000: In the rising period 2003-2008, all members of the triplet are clearly lower than in the period before 2000, indicating that the rise during 2003-2008 is less non-extensive in character than the rise before 2000. This is also indicated by the values of the $Q$–metrics, which are clearly lower, signifying a more efficient and normal market. This is evidence in support of the fact that the price rise in the period 2003-2008 does not have the characteristics of a pure stock-market bubble. Furthermore, during the 2008 crash period (2008-2009), which is not attributed to an inherent stock market crisis, the behavior of the $q$-triplet is again different from the crash period of 2000 as it shows the exact opposite trends: $q_{stat}$ rises instead of dropping, $q_{sen}$ drops instead of rising and $q_{rel}$ rises instead of dropping. As a consequence, $Q^{inv}$ rises during the 2008 crash instead of dropping, indicating that the non-extensivity increases after the crash, whereas in a pure stock market bubble non-extensivity is reduced after the crash. Notice also that in the rapidly rising period 2010-2020, the triplet shows mildly non-extensive characteristics, which again indicates a market rise characterized by *efficiency*, rather than a developing bubble.

Another interesting result is the negative width *W* and *B*-proxy value (within standard error) during the 2008 crash period. This is a consequence of the fact that $H_4 > H_1 > H_{0.1}$. This phenomenon was called '*reversal*' by in (Antoniades, 2020) and is a kind of anomalous scaling which happens when $H_4$ (which weighs the tails of the difference distribution) behaves more 'persistently' than the central points. This can occur when large consecutive daily % drops of a stock market index price are all in the same direction (which happens during a prolonged market crash, like the one in 2008). On the other hand, a market crash such as, for example, Black Monday in which a very large drop was immediately followed by a large rise (or vice-versa), has an *anti*-persistent effect on the high-$q$ value $H_q$, leading to $H_4 \ll H_1$ and a very large positive *W*.

*NIKKEI Stock Market*

**Table 3** shows the time-periods in which the NIKKEI stock market index was divided including the market events that mark the limits of each time period.

**Table 3:** Description of the break-down of NIKKEI close price timeseries in time periods used for the Tsallis statistics, GHE and $Q$-metrics calculations.

| Period No | Duration (Tr. Days) | | Start/End Dates | Related market event |
|---|---|---|---|---|
| | | | | |

| 1 | 1624 | Start | 6/1/1965 | Beginning of recorded data |
| --- | --- | --- | --- | --- |
| | | End | 22/12/1971 | (roughly) Beginning of Japanese economy bloom |
| 2 | 1881 | Start | 27/12/1971 | " |
| | | End | 10/1/1980 | (roughly) Beginning of Japanese 1990 bubble rising period |
| 3 | 1855 | Start | 11/1/1980 | " |
| | | End | 16/10/1987 | Just before 'Black Monday' event. |
| 4 | 529 | Start | 19/10/1987 | 'Black Monday' event. |
| | | End | 12/1/1990 | (roughly) End of Japanese 1990 bubble rising period |
| 5 | 616 | Start | 17/1/1990 | (roughly) Beginning of Japanese 1990 bubble dropping period |
| | | End | 20/8/1992 | (roughly) End of Japanese 1990 bubble dropping period |
| 6 | 1813 | Start | 21/8/1992 | (roughly) Beginning of Japanese economy stagnation period |
| | | End | 9/5/2000 | (roughly) End of 2000 dot.com bubble rising period |
| 7 | 707 | Start | 10/5/2000 | (roughly) Beginning of 2000 dot.com bubble crash |
| | | End | 9/5/2003 | (roughly) End of 2000 dot.com bubble crash |
| 8 | 1063 | Start | 10/5/2003 | (roughly) Beginning of mid-millennia's blooming period |
| | | End | 31/10/2007 | (roughly) End of mid-millennia's blooming period |
| 9 | 613 | Start | 1/11/2007 | (roughly) Beginning of 2008 US real estate related market crash |
| | | End | 20/5/2010 | (roughly) End of 2008 US real estate related market crash |
| 10 | 1290 | Start | 21/5/2010 | " |
| | | End | 20/8/2015 | Middle point of 'teens' decade. |
| 11 | 1077 | Start | 21/8/2015 | " |
| | | End | 13/2/2020 | End of recorded data |

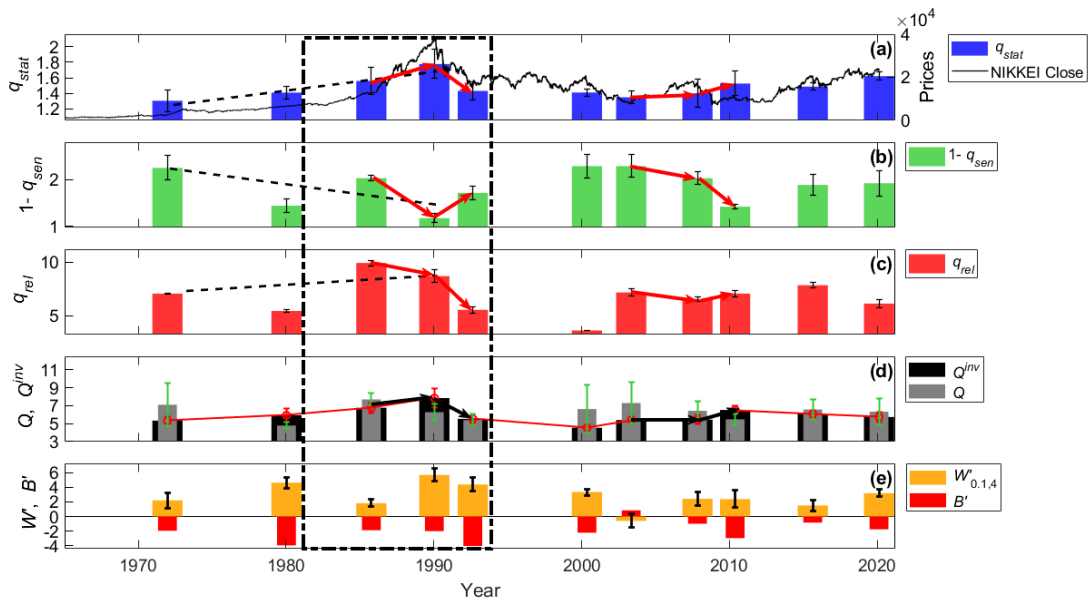

[15]

**Figure 2:** Time-dependent Tsallis $q$-triplet, $Q$-metrics, normalized GHE width $W'$ and normalized $B$-proxy $B'$ for the NIKKEI stock market daily closing price timeseries for the time-periods shown in Table 2. a) $q_{stat}$, b) $q_{sen}$ c) $q_{rel}$, d) $Q$ and $Q^{inv}$, (e) $W'_{0.1,4}, B'_{0.1,1}$. The striped box encloses two periods before the 1990 bubble break and the period after the break. The red arrows in (a), (b), (c) show the local trends in the $q$-triplet before and after the bubble break for the 1990 and the 2008 crises and the black arrows in (d) the respective trends in $Q^{inv}$. The solid red line in (d) is a line plot of metric $Q^{inv}$ and serves merely as a guide to the eye showing all the local trends in $Q^{inv}$. The striped black lines show linear least squares fits to the data before the year 2000 crisis for each index of the $q$-triplet.

**Figure 2** shows the results of the Tsallis $q$-triplet, the normalized GHE width $W'$, $B'$ and the $Q$-metrics for the NIKKEI index. We observe that the $q_{stat}$ index varies between $1.3 \pm 0.14$ and $1.88 \pm 0.18$. The maximum value of $q_{stat}$ (1.88) was observed in period 1980-1987 and is associated with the Japanese 1990 bubble rising period. The relative trend in all $q$-indices as well as in $W$ in this and the post bubble break period are very similar to the respective year 2000 dot.com bubble in S&P 500. Here too, we observe the same pattern: a rising trend in $q_{stat}$ before the 1990 bubble peak, followed by steep dropping trend after the bubble break; a strongly multifractal/multiscaling behavior followed by monofractal/uniscaling behavior (manifested by the steep increase in $q_{sen}$/steep decrease in $W$ respectively); a corresponding relative drop in the $Q^{inv}$ metric. In the next periods until 2008 (all of them within the Japanese economy stagnation period), we observe that $q_{stat}$ remains small and almost constant varying between 1.35 and 1.41. It shows a small rise before the 2000 dot.com bubble followed by a drop afterwards, while in the 2008 US real estate crises, it shows the reverse trend: it is smaller before the crash than after it. In the last period, 2010-2020 $q_{stat}$ rises again (just like in the respective period for S&P 500) up to a value of $1.62 \pm 0.05$ following the rising trend of the NIKKEI price. The period 1992-2008 (stagnation period) is characterized by market efficiency and this is manifested by all three $q$-indices, as well as $W$ and $B$: a small $q_{stat}$, a large $1 - q_{sen}$ (denoting monofractal or weak multifractal behavior), small $W, B$ (denoting uniscaling or weak multiscaling) and a small $q_{rel}$ denoting an increase of the effective number of degrees of freedom of the attractor of the underlying dynamics. This behavior is summarized by the relative values of the $Q$ and $Q^{inv}$ metric: $Q$ is high (affected by the large values of $q_{sen}$), whereas $Q^{inv}$ is very small. It is also very interesting to discuss the last period (2010-2020) in terms of the relative variations of the $q$-indices and $W$: the rising of $q_{stat}$ denotes a fattening of the price change distribution tails, whereas the moderate lowering of $q_{sen}$ relative to the stagnation period (and equivalently the small increase in $W', B'$) indicates that the market remains moderately multifractal/multiscaling during the NIKKEI price rise. It is also interesting that $q_{rel}$ remains at the same small level as the stagnation period, indicating that the underlying dynamics is still highly-dimensional. All this suggests that the market growth during 2010-2020 is on the one hand characterized by more non-extensive statistics relative to stochastic dynamics as far as the price change distribution is concerned, but on the other hand, in terms of multiscaling and relaxation of disturbances, it is much closer to stochastic behavior (efficient market) than the rising period of the 1990 bubble. This behavior is somewhat captured by the relative values of the $Q$ and $Q^{inv}$ metrics: $Q^{inv}$ has a relatively low value and $Q$ is slightly higher (the values are very similar to the period after the 1990 crash). These results indicate that for S&P 500 and NIKKEI (both major markets) during rising periods of a developing stock market bubble, show pronounced non-extensive statistics and a high value of the $Q^{inv}$ metric. In efficient market periods, $Q^{inv}$ is low and $Q$ high, usually higher than $Q^{inv}$. However, there can be periods of market growth, during which the market shows a clear rising trend but it is also efficient, meaning that the growth is not similar to that of a developing bubble. In such a case, $q_{stat}$ may be higher than expected for a completely efficient market, (due to more pronounced tails in the price difference distribution), but $q_{rel}$ still remains low and $1 - q_{sen}$ high (or equivalently, $W$ and $B$ low), indicating high-dimensionality and uniscaling behavior, which are both characteristics of an efficient market. Another interesting result is that $W', B'$, although they have the same relative magnitude in most periods, they differ significantly in the period just before the 1990 crash. We checked that this is due to the trading period immediately after 'Black Monday' (Oct. 18-25, 1987) which is contained in the period no 4 (19/10/1987-12/1/1990). This period contains a large number of extreme tail events which affect *W* much more than *B*.

*DAX Stock Market*

After examining an American and Asian major index, we present the results of the analysis for DAX, a European major index. Table 4 shows the periods for which the *q*-triplet, *Q*-metrics, *W* and *B* were computed for the DAX index. As in a previous table, shaded cells correspond to dates with no particular known event, and were chosen in order to have approximately an equal number of trading days in neighboring time periods.

Table 4: Description of the break-down of DAX close price timeseries in time periods used for the Tsallis statistics, GHE and *Q*-metrics calculations.

| Period No | Duration (Tr. Days) | | Start/End Dates | Related market event |
|---|---|---|---|---|
| 1 | 1150 | Start | 30/12/1987 | Beginning of recorded data |
| | | End | 5/10/92 | (roughly) Beginning of 90's market blooming period |
| 2 | 1116 | Start | 6/10/92 | " |
| | | End | 7/5/97 | 4 months before Asian market crisis |
| 3 | 734 | Start | 8/5/1997 | " |
| | | End | 5/5/2000 | (roughly) End of the year 2000 dot.com bubble rising period. |
| 4 | 750 | Start | 6/5/2000 | (roughly) Beginning of the year 2000 dot.com bubble crash. |
| | | End | 10/5/2003 | (roughly) End of the year 2000 dot.com bubble crash |
| 5 | 1139 | Start | 12/5/2003 | (roughly) Beginning of mid-millennia's rising period |
| | | End | 31/10/2007 | (roughly) End of mid-millennia's rising period |
| 6 | 382 | Start | 1/11/2007 | (roughly) Beginning of 2008 US real-estate crisis crash. |
| | | End | 9/5/2009 | (roughly) End of 2008 US real-estate crisis crash. |
| 7 | 1384 | Start | 10/5/2009 | " |
| | | End | 15/10/2014 | |
| 8 | 1339 | Start | 16/10/2014 | |
| | | End | 13/2/2020 | (roughly) End of recorded data |

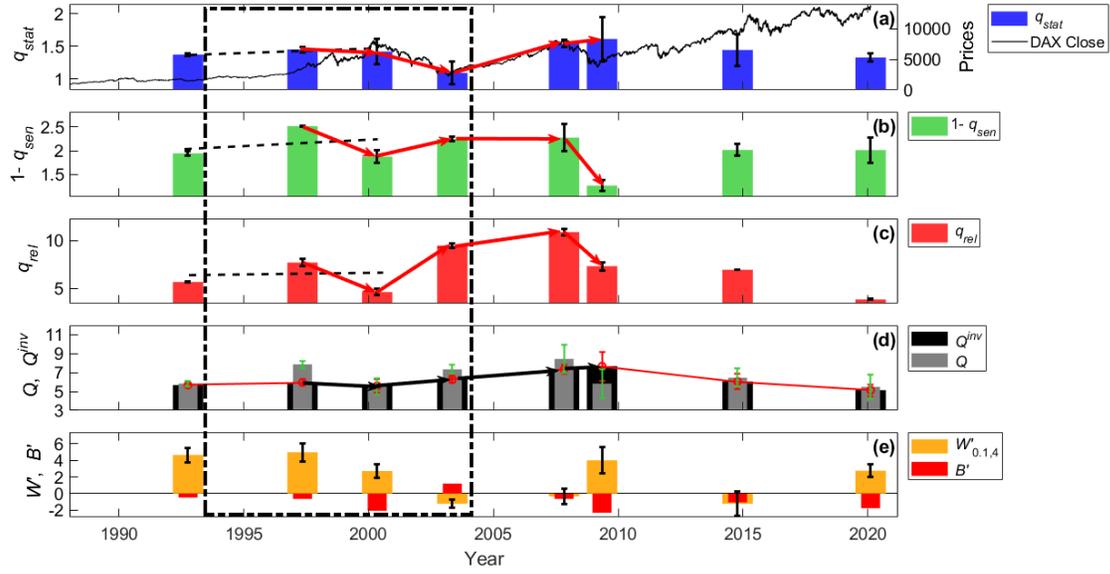

Figure 3: Time-dependent Tsallis $q$-triplet, $Q$-metrics, normalized GHE width $W'$ and normalized $B$-proxy $B'$ for the DAX stock market daily closing price timeseries for the time-periods shown in Table 2. a) $q_{stat}$, b) $q_{sen}$ c) $q_{rel}$, d) $Q$ and $Q^{inv}$, (e) $W'_{0.1,4}, B'_{0.1,1}$. The striped box encloses two periods before the 2000 bubble break and the period after the break. The red arrows in (a), (b), (c) show the local trends in the $q$-triplet before and after the bubble break for the 2000 and the 2008 crises and the black arrows in (d) the respective trends in $Q^{inv}$. The solid red line in (d) is a line plot of metric $Q^{inv}$ and serves merely as a guide to the eye showing all the local trends in $Q^{inv}$. The striped black lines show linear least squares fits to the data before the year 2000 crisis for each index of the $q$-triplet.

Figure 3 shows the non-extensive statistics, $W'$, $B'$ and $Q$-metrics results for the DAX index. We observe a similar pattern in the relative $q$-triplet and $Q$-metric values as with S&P 500 and NIKKEI before and after the 2000 dot.com bubble. Also, we notice that the results for DAX are more similar to NIKKEI than S&P 500, for this period, as the dot.com crisis presented an external influence on DAX and NIKKEI, not an 'internal affair' as for S&P 500. In general, DAX seems to be characterized by much more weakly multifractal/multiscaling dynamics compared to both NIKKEI and S&P 500, throughout its timeline. This is verified by the overall higher $1 - q_{sen}$ values and equivalently smaller $W', B'$ values that DAX has in all time periods. It is interesting to note that, for DAX too, the $Q^{inv}$ index displays the reverse behavior for the periods before and after the 2008 crisis compared to the periods before and after the 2000 crisis, as is also remarkably true for NIKKEI and S&P 500: $Q^{inv}$ rises after the crash of 2008, whereas it drops after the crash of 2000. However, $Q^{inv}$ still rises before the 2008 peak relative to the previous period for DAX, as well as the other markets, but this is mainly due to the rise in $q_{rel}$ since the $q_{stat}$ is low and $W, B$ too. This means that the rising trend in the period before 2008 was not due to a developing stock market bubble, but rather an efficient market rise due to the millennial global economic growth. The much unexpected (by stock markets) 2008 crash was due to an exogenous financial event and thus profoundly differs from the pure stock market bubbles of 2000. This is evidence that the $q$-triplet and the $Q^{inv}$ metric temporal evolution essentially allows us to distinguish between an endogenous and an exogenous stock-market crisis.

A final notable fact is that the normalized $B$-proxy value for periods no 2 and 3 shows an opposite trend from $W'_{0.1,4}$ in the same periods: $B'$ rises while $W'$ drops. The rise of $B'$ is consistent with the fact that $1 - q_{sen}$ drops going from period no 2 to no 3. The reader is reminded that a dropping $1 - q_{sen}$ and equivalently a rising $B'_{0.1,1}$ or a rising $W'_{0.1,4}$ signifies a transition to a more multiscaling state. However, the dropping of $W'_{0.1,4}$ seems to contradict this trend. This is an example of a situation where $W'_{0.1,4}$ and $B'_{0.1,1}$ convey different type of information regarding multiscaling, as they depend on a different range of values for the parameter $q$. The fact that $W'_{0.1,4}$ is much larger than $B'_{0.1,1}$ in the period 1987-1997 suggests that, on the one hand, there are statistically many more extreme tail events than there exist, on the average, for a random surrogate index, while on the other hand, the small and medium-sized events occurred roughly at the same proportion with the random surrogate.

[18]

*LSE Stock Market*

Another major European market is the London Stock Exchange (LSE) to which we applied our analysis for the period 2001-2020. Table 5 shows the breakdown in time periods in which we computed the $q$-triplet, $Q$-metrics, $W$ and $B$ of the LSE close price timeseries. As in previous tables, shaded cells correspond to dates with no particular known event, and were chosen so that neighboring time periods contain approximately the same number of trading days.

Table 5: Description of the break-down of LSE close price timeseries in time periods used for the Tsallis statistics, GHE and $Q$-metrics calculations.

| Period No | Duration (Tr. Days) | | Start/End Dates | Related market event |
|---|---|---|---|---|
| 1 | 885 | Start | 23/7/2001 | Begin of recorded data. |
| | | End | 10/12/2004 | |
| 2 | 734 | Start | 11/12/2004 | |
| | | End | 30/10/2007 | roughly) End of mid-millennia's rising period |
| 3 | 377 | Start | 31/10/2007 | (roughly) Beginning of the LSE 2007 crash. |
| | | End | 10/5/2009 | (roughly) End of 2007 LSE crash. |
| 4 | 768 | Start | 11/5/2009 | " |
| | | End | 31/5/2012 | |
| 5 | 780 | Start | 1/6/2012 | |
| | | End | 15/7/2015 | |
| 6 | 803 | Start | 16/7/2015 | |
| | | End | 21/9/2018 | (roughly) Apple Corp. related crash. |
| 7 | 353 | Start | 22/9/2018 | " |
| | | End | 14/2/2020 | End of recorded data |

[19]

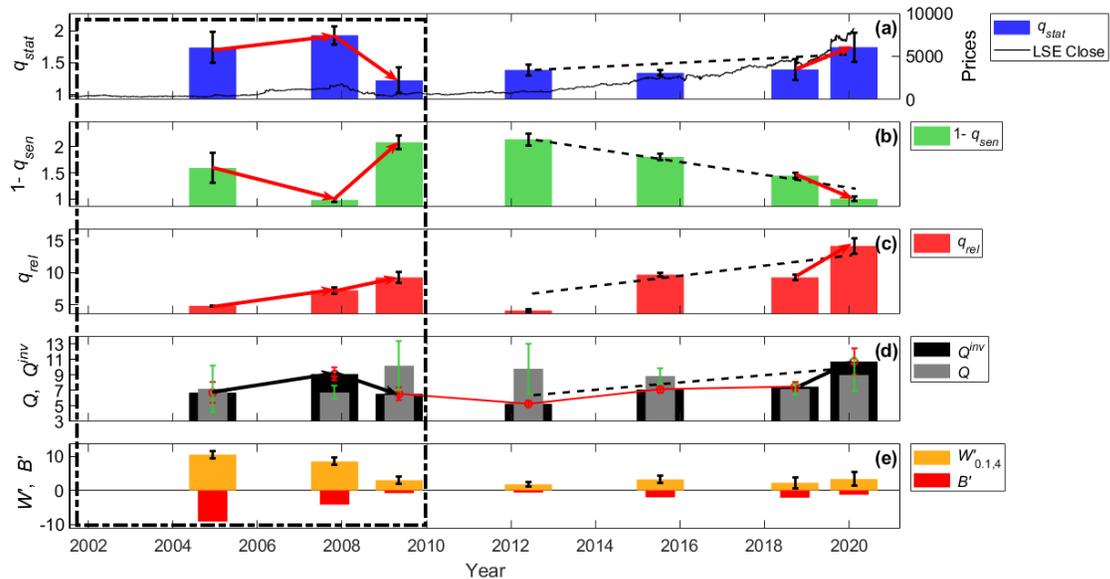

Figure 4: Time-dependent Tsallis $q$-triplet, $Q$-metrics, normalized GHE width $W'$ and normalized $B$-proxy $B'$ for the LSE stock market daily closing price timeseries for the time-periods shown in Table 2. a) $q_{stat}$, b) $q_{sen}$ c) $q_{rel}$, d) $Q$ and $Q^{inv}$, (e) $W'_{0.1,4}, B'_{0.1,1}$. The striped box encloses two periods before the 2008 bubble break and the period after the break. The red arrows in (a), (b), (c) show the local trends in the $q$-triplet before and after the bubble break for the 2008 crisis as well as for the last two periods before Feb.14, 2020 which was the last recorded date of the data series. The black arrows in (d) show the respective trends in $Q^{inv}$. The solid red line in (d) is a line plot of metric $Q^{inv}$ and serves merely as a guide to the eye showing all the local trends in $Q^{inv}$. The striped black lines show linear least squares fits to the data in the period 2010-2020 for each index of the $q$-triplet.

Figure 4 shows the non-extensive statistics, $W', B'$ and the respective $Q$-metrics results for the LSE index. As can be observed, $q_{stat}$ varies between 1.22 ± 0.14 and 1.93 ± 0.20. The maximum value of $q_{stat}$ (1.93) is observed in the period 2001-2007 and is connected to the period prior to the 2008 US real-estate crisis related crash. During the crash, we observe a notable reduction of the $q_{stat}$ index to its lowest value of 1.22. This is similar behavior to the one we observed for the dot.com bubble for S&P 500 and the Japanese bubble in 1990 for NIKKEI. It also agrees with the rest of the $q$-triplet indices as well as the $Q$-metrics. We conclude that the 2008 crash for LSE does not present the same non-extensive statistics evolution and scaling evolution as the 2008 crash for NIKKEI, DAX and S&P 500 and has probably different causes. In fact, the rise of LSE in the period before 2007, shows characteristics of a developing pure stock market bubble, the break of which coincided with the US real estate crisis (and was probably triggered by it). This observation strengthens further the assumption that there is a direct connection between a developing stock market bubble and the observation of high-$q_{stat}$/low $(1 − q_{sen})$/high $q_{rel}$/high $Q^{inv}$/high $W'$ in that time period which is similar to the non-extensive statistics and scaling in before stock market bubbles in the other stock indices.

Finally, we observe than in the period 2009-2020 LSE shows a rising period of rapid and prolonged market growth characterized by a more efficient market (indicated by the smaller $q_{stat}$ and somewhat smaller $W'$ relative to the prior to 2007 period), but it still shows a rather large and rising $q_{rel}$ as well as a rising trend in $q_{stat}$ and dropping trend in $1 − q_{sen}$. Therefore, this growth contains more 'bubble'-like characteristics relative to the same period for the other major indices (DAX, NIKKEI and S&P 500) and could signify that LSE is currently within a period of bubble development.

[20]

In order to depict the temporal trends in the $q$-triplet and how they compare among the four markets under study, specifically during critical periods (stock market bubbles), firstly, we computed the slopes (i) between the period *just after* the bubble break and the period *just before* the break and (ii) between the period just *before* the bubble break and the period before that. These represent the *local* trends in the periods just before and just after a bubble break that reveal relative temporal patterns among the indices of the q-triplet that are specific to the periods just before and after a critical event. We used the year 2000 bubble for S&P 500 and DAX, year 1990 bubble for NIKKEI and year 2008 for LSE. Secondly, we performed least-square linear fits using $q$-triplet data from a longer period before the particular critical events and recorded the slopes of the fitted lines. These represent more long-term trends before the bubble break and reveal respective long-term temporal patterns that together with the local trends may represent warning signals for a bubble under development. (The fitted lines are shown with dashed lines for each market in Figure 1 - Figure 4 respectively and the periods used for the fits are apparent there). LSE was excluded from the latter analysis because we had only two periods before the 2008 bubble as there was no data available prior to 2001 for LSE. In order to enable comparisons among slopes for the different indices of the triplet and the $Q^{inv}$ metric, we divided the slopes by $(x_{max} - x_{min})$, where $x$ is the respective index $q_{stat}$, $(1 - q_{sen})$, $q_{rel}$ or $Q^{inv}$, and subscripts *max*, *min* denote the maximum and minimum value of the respective quantity for all periods. Performing this normalization, we brought the slopes of all $q$-triplet indices and markets to the same scale. In Figure 5 we plot the normalized slopes for each member of $q$-triplet and $Q^{inv}$. Figure 5a contains a bar chart that shows the *local* trends *before* (boldface bars) and *after* (transparent bars) the bubble breaks and Figure 5b a bar chart with the *long-term* trends *before* the bubble. Inspecting the local trends, the similarities in the before-after crisis trend patterns for index $q_{stat}$, $(1 - q_{sen})$, as well as $Q^{inv}$ among all four markets is apparent: There is a rising trend for $q_{stat}$ just before the bubble break and a dropping trend just after. The exception is DAX before crisis trend which is practically neutral. The rise in $q_{stat}$ is connected to an increased departure from Gaussian shape of the price change distribution with a fattening of tails due to an increase in the number of big and medium sized changes. After the break the price change distribution gradually returns to a more Gaussian shape. From Figure 5b we also see a more long-term rising trend in $q_{stat}$ in all three markets before the bubble break. Regarding $(1 - q_{sen})$, there is a clear dropping trend which is connected to an increase in multifractality before the break, followed by a rising trend, which suggests that the series become more unifractal after the bubble break, *i.e.* the market returns to efficiency. There is a more long-term dropping trend in $(1 - q_{sen})$ for S&P 500 and NIKKEI. DAX shows a slightly dropping long-term (almost neutral) trend before the 2000 bubble, but it is also interesting to note that all three markets depict the following pattern in $(1 - q_{sen})$: Firstly, a rise, then a drop in the period just before the bubble breaks. This suggests that in the initial stage of bubble development the timeseries is more unifractal, but during the period just before bubble break multifractality increases while it decreases after the break. Finally, $q_{rel}$ shows a dropping local trend before the bubble break in S&P 500, NIKKEI and DAX 2000 crisis, and a rising trend in LSE 2008 crisis, while there are dropping trends after the break with increased magnitude for S&P 500 and NIKKEI, but rising trends for DAX and LSE. Long-term trends in $q_{rel}$ are clearly rising before the bubble break for S&P 500 and NIKKEI while slightly dropping for DAX. There seems to be a pattern here too: $q_{rel}$ increases during bubble development, but stabilizes or slightly decreases in the period just before the bubble break. Since $q_{rel}$ is connected with the dimensionality of the dynamical system, it could be connected with herding behavior, which increases (in general) during bubble development, but stabilizes or decreases just before the bubble break, when different opinions of traders about the future of the market may be the cause of a sudden increase in the dimensionality of trading dynamics.

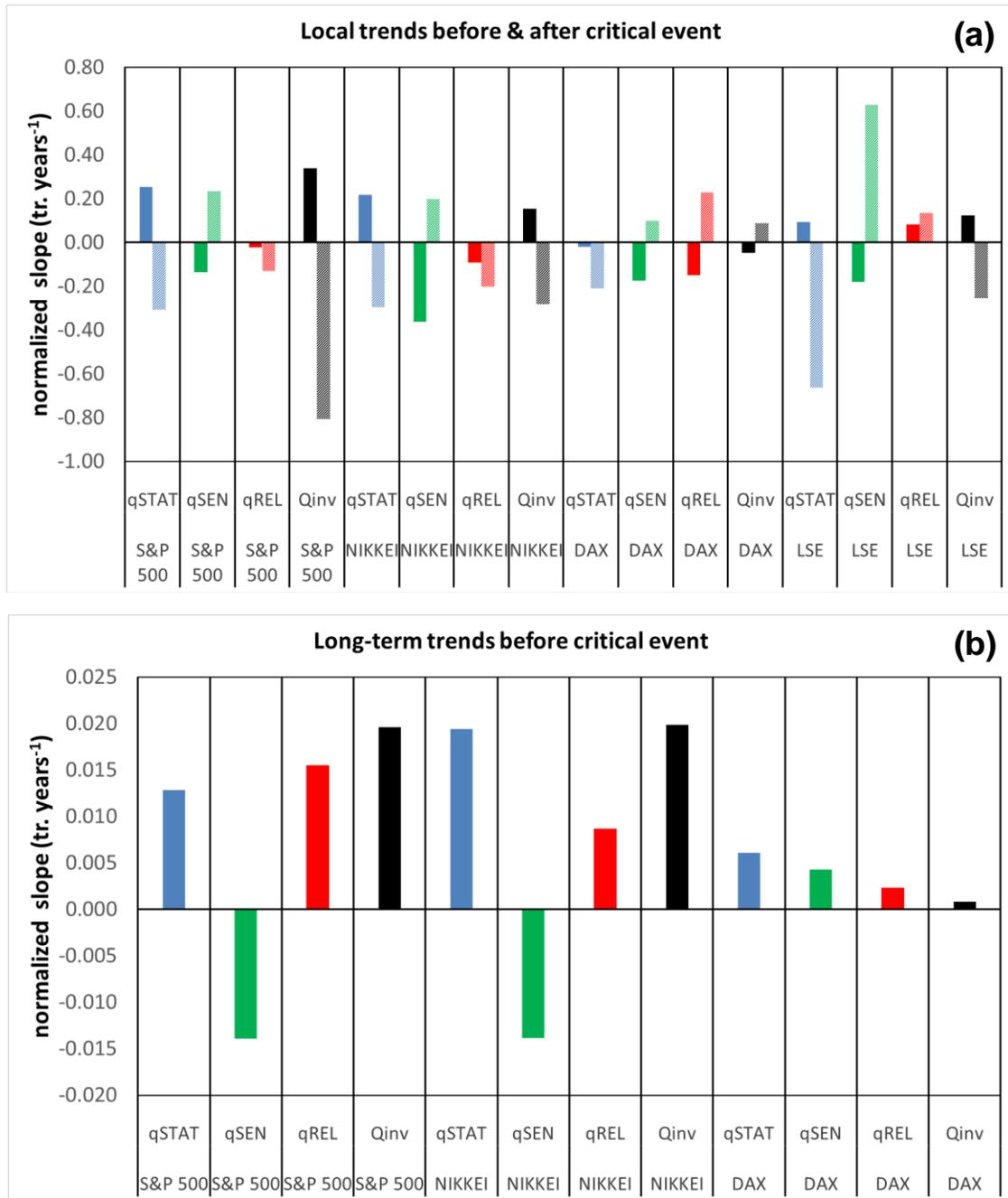

Figure 5: Comparison between the temporal trends of the $q$-triplet and $Q^{inv}$ metric *before* and *after* a market crisis for the four markets considered. For each index in the $q$-triplet and for $Q^{inv}$ we plot: (a) (Local trend) the normalized difference between (i) the period *just after* the bubble break and the period *just before* the break and (ii) the period just *before* the bubble break and the period before that; (b) (long-range trend): the normalized slope obtained by a least-squares linear fit *before* the bubble break as shown in Figure 1-Figure 4. The year 2000 dot.com crisis is chosen for SP&500 and DAX, the year 1990 crisis for NIKKEI and the year 2008 crisis for LSE. The color coding used for each triplet index $q_{stat}$, $q_{sen}$, $q_{rel}$ and $Q^{inv}$, blue, green, red, black in respect, is the same as in Figure 1 - Figure 4. Boldface bars correspond to values *before* the crisis, and transparent colors to the values *after* the crisis.

Finally, in Figure 6 we show a scatter plot of $(1 - q_{sen})^{-1}$ vs. (a) $W'$ and (b) $-B'$ for all time periods and market indices. As already mentioned in equation (18) $(1 - q_{sen})^{-1}$ is proportional to the multifractal spectrum width $\Delta\alpha$ while $W'$ and (b) $B'$ are also measures of the multifractality of the timeseries coming from the GHE spectrum. It is interesting to see the correlations among these three quantities in the various time periods for the various markets and whether there exist patterns that distinguish between periods before a crisis and periods after as well as patterns that distinguish between endogenous crises and exogenous crises. As can be seen most of the points cluster in along the striped black line which is a linear least squares fit to the data points enclosed in the ellipse. These points show that there is a strong correlation between $(1 - q_{sen})^{-1}$ and $W'$ as well as $B'$, as expected from theory. However, there are other data points that are clear outliers, falling far away from the region of correlation. We have labeled all these outliers by the period they correspond to, as shown by the colored dotted arrows. Almost all these outliers correspond to the periods just before or just after a bubble break, *a fact that suggests that market period around a crisis show anomalous correlation behavior between the values of $(1 - q_{sen})^{-1}$ and either $W'$ or $B'$*. Another interesting fact is that the rising periods just before the pure stock market bubbles (2000 for S&P 500, DAX, 1990 for NIKKEI and 2008 for LSE) all lie on or above the line of regression, or equivalently show a very large $(1 - q_{sen})^{-1}$ and from a very moderate value of $W'$ (S&P 500, DAX) up to a large (NIKKEI) or very large value of $W'$. For the plot vs. $-B'$, the same general picture is observed with NIKKEI showing relatively lower $-B'$ for the 1990 bubble rising period. The differences in $(1 - q_{sen})^{-1}$ among different markets during this period are quite small, whereas the respective differences in $W'$ or $B'$ can be large. The exact same observation can be made for the period just after the 2000(1990 NIKKEI, 2008 LSE) crash: the data points are clustered under the correlation line at low values *and very similar* values of in $(1 - q_{sen})^{-1}$ (all within ~0.45(DAX) - ~0.6(LSE), whereas values of $W'$ or $B'$ in the crash period vary greatly, from ~ -1.3(DAX) up to ~10.5(LSE). This suggests that *the three multifractality indicators convey different information in critical periods and that the correlation among them is broken*. Notice also that the year 2008 exogenous crisis for S&P 500, NIKKEI and DAX has the reverse trend in the periods before and after the crisis than the purely endogenous crisis of 2000 and 1990: The 2008 crash period has large $(1 - q_{sen})^{-1}$ while the 2008 rise period has low $(1 - q_{sen})^{-1}$. This distinguishes between these types of crises. To make this difference clearer to the eye we marked the transition from the periods just before a crisis to just after with arrows: the color of the arrows correspond to the market index as shown by the legend. Bold arrow lines correspond to the considered pure stock market bubbles and striped arrow lines to the 2008 exogenous crisis. As can be seen from Figure 6a the transition for a pure stock market bubble is marked by a decrease in both $(1 - q_{sen})^{-1}$ and $W'$ but the magnitude of the decrease in each of the two indicators varies among different markets: for example for DAX, the decrease in $W'$ is large and in $(1 - q_{sen})^{-1}$ very small, whereas for S&P 500 vice-versa and for LSE both decreases are similar. As can be seen from Figure 6a, on the other hand, the transition in a pure stock-market bubble is again towards lower values of $(1 - q_{sen})^{-1}$ for all markets and lower $-B'$ for S&P 500, DAX and LSE whereas NIKKEI moves to a larger $-B'$. It is also interesting that, a comparison between transitions of the two types of crises for the same market, shows that they always occur in the exact opposite direction in both $(1 - q_{sen})^{-1}$ *and* $W'$ or $-B'$ (with the exception of NIKKEI for which the 2008 crisis transition showed almost no change in $W'$).

[23]

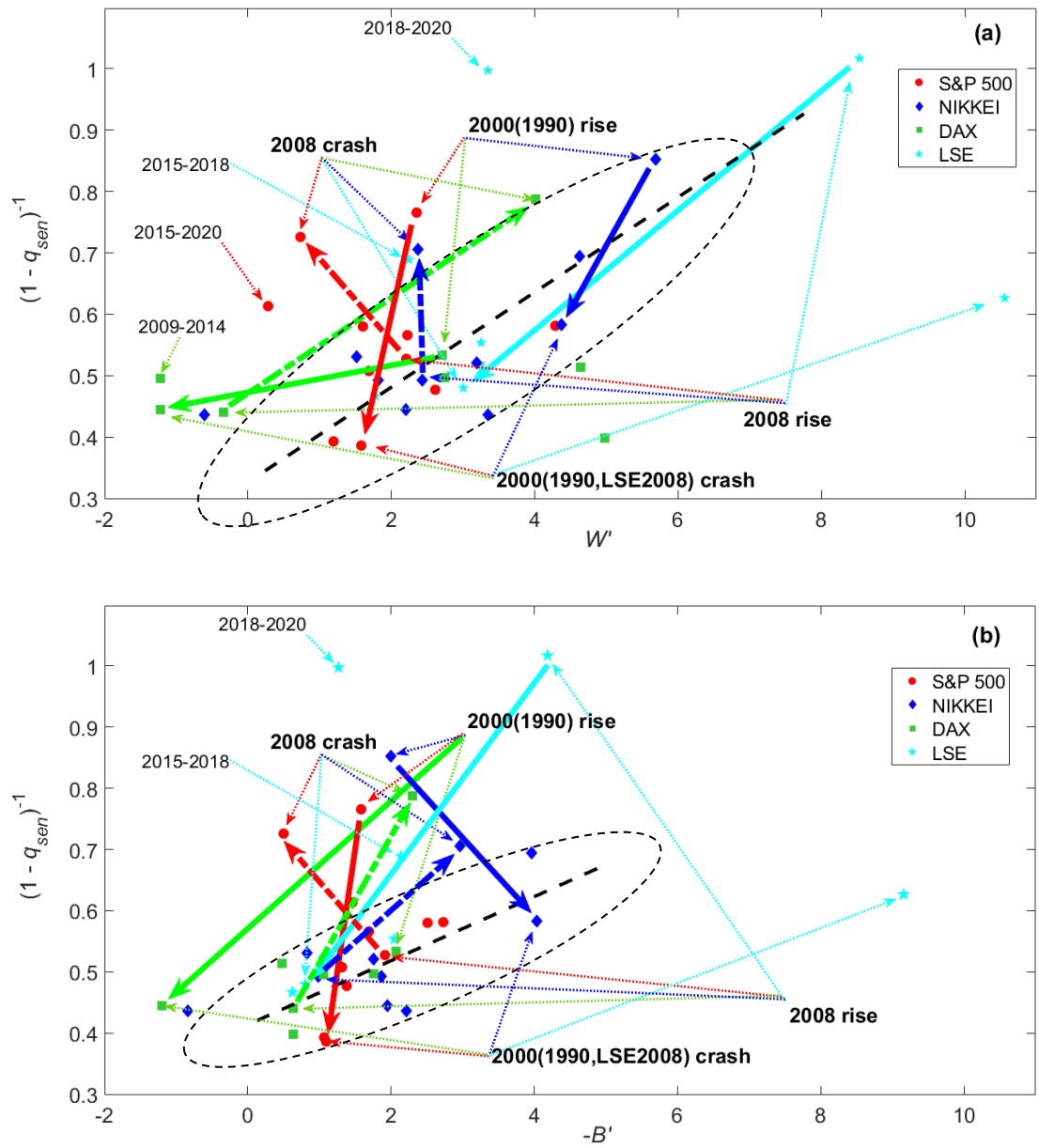

Figure 6: Correlation of $(1 - q_{sen})^{-1}$ vs. (a) $W'$ and (b) $B'$ for all time periods and market indices. The labels correspond to various market periods. Bolded labels correspond to the periods immediately before or after a crisis event. Bold solid arrows color coded per market index show the direction from the point of a market bubble peak to a bubble crash period (1990 for NIKKEI, 2000 for S&P 500, DAX and 2008 for LSE). Bold dashed arrows show the respective market direction for the 2008 real-estate related crisis for S&P 500, NIKKEI and DAX. The dashed straight black lines correspond to least-squares linear fits to the data point contained in the striped ellipses.

[24]

Table 6: *q*-triplet, *W*, *B* and *Q*-metric values per time period for each stock market index. Numbers in parentheses represent standard errors.

|  | **Period** | $q_{stat}$ | $1 - q_{sen}$ | $q_{rel}$ | $W'$ | $B'$ | $Q$ | $Q^{inv}$ |
|---|---|---|---|---|---|---|---|---|
| **S&P500** | 21/1/55-29/3/63 | 1.40 (0.08) | 1.72 (0.15) | 5.67 ( 0.2) | 4.29 (0.62) | -2.73 | 6.23 (0.77) | 6.56 (0.59) |
|  | 30/3/63-28/5/71 | 1.42 (0.08) | 1.77 (0.26) | 4.3 (0.18) | 2.22 (0.91) | -1.69 | 6.10 (1.44) | 6.24 ( 0.9) |
|  | 1/6/71-31/7/79 | 1.13 (0.06) | 2.55 (0.08) | 4.36 (0.11) | 1.19 (0.73) | -1.07 | 8.64 (1.54) | 4.89 (0.21) |
|  | 1/8/79-1/10/87 | 1.63 (0.14) | 1.97 (0.14) | 8.58 (0.34) | 1.69 (0.62) | -1.32 | 8.17 (1.08) | 7.41 (0.65) |
|  | 2/10/87-14/12/94 | 1.33 (0.05) | 2.10 (0.29) | 11.99 ( 0.1) | 2.61 (1.64) | -1.38 | 9.22 (2.16) | 8.02 (0.43) |
|  | 15/12/94-10/6/99 | 1.84 (0.19) | 1.31 (0.09) | 11.2 (0.45) | 2.35 (1.13) | -1.58 | 8.20 (1.07) | 9.62 (1.07) |
|  | 11/6/99-10/9/03 | 1.26 (0.10) | 2.59 (0.17) | 6.92 (0.19) | 1.58 (0.70) | -1.10 | 9.42 (2.73) | 5.8 (0.35) |
|  | 11/9/03-31/10/07 | 1.25 (0.07) | 1.90 (0.22) | 7.25 (0.15) | 2.21 (1.03) | -1.91 | 6.92 (1.47) | 6.46 (0.57) |
|  | 1/11/07-5/3/09 | 1.52 (0.16) | 1.38 (0.13) | 8.81 (0.22) | 0.73 (0.98) | -0.50 | 6.74 (0.69) | 8.25 (0.99) |
|  | 6/3/09-21/8/15 | 1.47 (0.09) | 1.72 (0.09) | 11.31 (0.22) | 1.6 (0.65) | -2.51 | 8.21 (0.51) | 8.33 (0.43) |
|  | 22/8/15-13/2/20 | 1.76 (0.23) | 1.63 (0.16) | 9.2 (0.25) | 0.28 (0.69) | -1.46 | 7.78 (1.23) | 8.25 (1.14) |
| **NIK-KEI** | 6/1/65-22/12/71 | 1.3 (0.14) | 2.25 (0.26) | 7.06 (0.03) | 2.21 (1.07) | -1.95 | 7.10 (2.38) | 5.33 ( 0.4) |
|  | 27/12/71-10/1/80 | 1.41 (0.08) | 1.44 (0.14) | 5.44 (0.15) | 4.63 (0.72) | -3.96 | 4.79 (0.37) | 5.94 (0.76) |
|  | 11/1/80-9/10/85 | 1.56 (0.17) | 2.03 (0.06) | 9.93 (0.26) | 1.81 (0.49) | -1.87 | 7.71 (0.66) | 6.78 (0.54) |
|  | 14/10/85-12/1/90 | 1.78 (0.19) | 1.17 (0.09) | 8.69 (0.59) | 5.69 (0.85) | -2.00 | 6.29 (0.94) | 7.85 (1.05) |
|  | 17/1/90-20/8/92 | 1.43 (0.12) | 1.72 (0.14) | 5.55 (0.29) | 4.38 (0.93) | -4.04 | 5.40 (0.67) | 5.54 (0.51) |
|  | 21/8/92-9/5/00 | 1.41 (0.05) | 2.29 (0.26) | 3.62 (0.02) | 3.35 (0.43) | -2.22 | 6.62 (2.67) | 4.54 (0.31) |
|  | 10/5/00-9/5/03 | 1.35 (0.08) | 2.29 (0.24) | 7.17 (0.32) | -0.6 (0.92) | 0.84 | 7.30 (2.30) | 5.40 (0.37) |
|  | 10/5/03-31/10/07 | 1.4 (0.18) | 2.03 (0.14) | 6.56 (0.18) | 2.43 (0.97) | -0.98 | 6.43 (1.04) | 5.45 (0.51) |
|  | 1/11/07-20/5/10 | 1.53 (0.16) | 1.42 (0.05) | 7.06 (0.28) | 2.37 (1.18) | -2.97 | 5.49 (0.57) | 6.53 ( 0.5) |
|  | 21/5/10-20/8/15 | 1.49 (0.05) | 1.88 (0.22) | 7.85 (0.23) | 1.51 (0.73) | -0.83 | 6.57 (1.09) | 6.09 (0.43) |
|  | 21/8/15-13/2/20 | 1.62 (0.05) | 1.92 (0.27) | 6.13 (0.37) | 3.19 ( 0.5) | -1.75 | 6.34 ( 1.5) | 5.74 (0.52) |
| **DAX** | 30/12/87-5/10/92 | 1.38 (0.02) | 1.94 (0.05) | 5.67 (0.06) | 4.64 (0.87) | -0.48 | 5.83 (0.25) | 5.68 (0.16) |
|  | 6/10/92-7/5/97 | 1.46 (0.04) | 2.52 (0.01) | 7.71 (0.35) | 4.98 (1.08) | -0.64 | 7.87 (0.39) | 5.96 (0.30) |
|  | 8/5/97-5/5/00 | 1.42 (0.19) | 1.88 (0.13) | 4.62 (0.33) | 2.71 (0.82) | -2.07 | 5.5 (0.86) | 5.60 (0.72) |
|  | 6/5/00-10/5/03 | 1.10 (0.17) | 2.25 (0.05) | 9.47 (0.21) | -1.22 (0.48) | 1.20 | 7.35 (0.50) | 6.26 (0.38) |
|  | 12/5/03-31/10/07 | 1.54 (0.05) | 2.27 (0.29) | 10.9 (0.30) | -0.34 (0.91) | -0.64 | 8.44 (1.50) | 7.32 (0.49) |
|  | 1/11/07-9/5/09 | 1.62 (0.34) | 1.27 (0.12) | 7.33 (0.44) | 4.01 (1.55) | -2.30 | 5.84 (1.55) | 7.67 (1.52) |
|  | 10/5/09-15/10/14 | 1.45 (0.24) | 2.02 (0.13) | 6.95 (0.03) | -1.22 (1.45) | -1.06 | 6.46 (0.95) | 6.06 (0.78) |
|  | 16/10/14-13/2/20 | 1.33 (0.06) | 2.01 (0.26) | 3.85 (0.06) | 2.75 (0.75) | -1.75 | 5.49 (1.30) | 5.14 (0.61) |
| **LSE** | 23/7/01-10/12/04 | 1.74 (0.24) | 1.60 (0.29) | 4.85 (0.08) | 10.56 (1.03) | -9.16 | 7.21 (3.03) | 6.71 (1.34) |
|  | 11/12/04-30/10/07 | 1.93 (0.14) | 0.98 (0.04) | 7.25 (0.52) | 8.53 (1.08) | -4.19 | 6.73 (0.92) | 9.11 (0.81) |
|  | 31/10/07-10/5/09 | 1.22 (0.20) | 2.08 (0.13) | 9.26 (0.78) | 3.00 (0.99) | -0.80 | 10.18(3.17) | 6.57 (0.86) |
|  | 11/5/09-31/5/12 | 1.39 (0.08) | 2.14 (0.12) | 4.12 (0.19) | 1.78 (0.71) | -0.62 | 9.78 (3.18) | 5.28 (0.34) |
|  | 1/6/12-15/7/15 | 1.34 (0.04) | 1.81 (0.06) | 9.7 (0.27) | 3.25 (1.02) | -2.04 | 8.85 (1.04) | 7.12 (0.43) |
|  | 16/7/15-21/9/18 | 1.39 (0.16) | 1.45 (0.05) | 9.26 (0.36) | 2.25 (1.57) | -2.14 | 7.18 (0.70) | 7.49 (0.60) |
|  | 22/9/18-14/2/20 | 1.74 (0.22) | 1.00 (0.04) | 14.16 (1.15) | 3.35 (2.00) | -1.27 | 8.98 (2.07) | 10.72 (1.71) |

## DISCUSSION

In this section we summarize the main results of the previous section and discuss their significance. Firstly, in all time periods and all stock markets, non-extensive statistics are observed within standard error, for all indices of the *q*-triplet, showing a clear departure from stochastic behavior. This fact agrees with other studies. Values of $1 - q_{sen}$ and similarly *W* and *B* (the width and depth of the GHE spectrum) indicate that for some periods there is clear multifractal or multiscaling behavior whereas for some others there is strong monofractal or uniscaling behavior. The transition from multiscaling to uniscaling occurs after a pure (endogenous) stock market bubble break (such as the Japanese year 1990 bubble and the year 2000 dot.com bubble), but not necessarily after an exogenous crisis (such as the 2008 US real estate triggered market crash). In the rising period before an endogenous stock market bubble (such as the Japanese year 1990 bubble and the year 2000 dot.com bubble), $q_{stat}$ shows a clear rising trend, both locally (~3 year period before bubble peak) and long-term (in a period 10-20 years depending on the market) peaking in the period just before the bubble break, indicating that the price difference distributions in such periods are characterized by more extreme events (fatter tails) than in other periods. This is a very similar finding to (Antoniades, 2020). One physical explanation of this fact (as was also argued in Antoniades, 2020) is that in the rising period of a developing bubble, there are several market traders with different strategies involved and there is also a large degree of trader 'nervousness' which is a result of the 'gut-feeling' many of them have about a coming crash. This causes *large drops usually followed by equally large rises* in the next trading day and, therefore, an increased frequency of tail events.

[25]

Moreover, in the rising period just before an endogenous stock market bubble, $1 - q_{sen}$ shows a clear dropping *long-term* trend, indicating increased multifractality while approaching the bubble break, which turns into a very strong dropping local trend in the period of the last ~3 years before the peak. In the period before the last ~3 years, while the market is already on the rise for the bubble, $1 - q_{sen}$ increases (or equivalently $(1 - q_{sen})^{-1}$ decreases), *i.e.* the time series appears to become more unifractal. This observation is consistent for all markets. In other words, in the development period of the bubble, the observed pattern is that the time series becomes first more unifractal then strongly multifractal, just before the bubble break. The fact that the series becomes more multifractal in the development period of the bubble is generally in accordance to $W'$ and $B'$ values that also indicate increased multiscaling, but the variability of $W'$ among the four markets is larger than the variability in $1 - q_{sen}$. Also, the local trend pattern of $W'$ and $B'$ in the last two periods before the break is not necessarily consistent with the pattern observed for $1 - q_{sen}$. For example, in NIKKEI, $W'$ drops then rises before the 1990 break, but not in S&P 500 and DAX. In DAX, we see this first-drop-then-rise pattern in $B'$ instead. In LSE, on the other hand, a slight drop in $W'$ and a clear drop in $B'$ is observed in the period just before the 2008 bubble break, which is opposite to the fact that $(1 - q_{sen})^{-1}$ rises then. This indicates that the behavior of the 2008 bubble in LSE lies somewhere between an endogenous market crisis and an exogenous one, such as what the 2008 crisis was for the major markets. The same 'nervous' market behavior and the fattening of price difference distribution tails is what causes the increased multiscaling combined with the fact that large price changes to one direction are most likely followed by large price changes in the opposite direction, which results in large-$q$ GHE's displaying a large drop (anti-persistence) relative to the small-$q$ GHE's and thus an increase in $W'$, but not necessarily $B'_{0.1,1}$ (see relative discussion in (Antoniades 2020)) in some cases. In other cases the reverse happens. We can conclude that $1 - q_{sen}$ is probably a more consistent across markets indicator of scaling during the development of a stock market bubble than $W'$ and $B'$ although the latter two capture different aspects of scaling, by weighing small and large price changes differently. The information revealed by Figure 6 supports the fact that all three indicators should probably be in order to obtain a more precise fingerprint of scaling and its temporal variation across markets.

Regarding the third index of the $q$-triplet, $q_{rel}$, we observe a *long-term* rising trend during the rising period before an endogenous stock market bubble indicating a clear departure from stochastic behavior and indicating the existence of a relatively low-dimensional attractor in phase space for system dynamics. This is a very interesting result that conveys different kind of information than the other members of the triplet and the multiscaling measures, *W* and *B*. Low dimensional dynamics could be a physical manifestation of '*herding*' behavior in trading during such periods, something that is captured by very large $q_{rel}$ values. However, the *local* trend just before the bubble break appears slightly reversed for S&P 500, NIKKEI and DAX as $q_{rel}$ slightly drops. For LSE, it rises but only slightly. This indicates that while the bubble peak approaches and anticipation of a possible breakdown increases among traders, the dimensionality of the dynamics slightly increases (or at least stabilizes), in other words 'herding' behavior lessens or stabilizes.

In the period after the break of an endogenous bubble, the $q$-triplet trends are reversed: $q_{stat}$ appears to drop (indicating the 'Gaussianization' of the price change distributions), $1 - q_{sen}$ rises (indicating the decrease of multifractality) and $W$ drops (indicating uniscaling behavior). Also, $q_{rel}$ drops to a value closer to 1, indicating the increase of the number of degrees of freedom available to the dynamics, and thus less coordinated dynamics (in other words, less 'herding').

In the periods before and after an exogenous market crisis (such as 2008 US real estate crisis), a different temporal evolution of trends is observed: $q_{stat}$ increases in the price falling period, and $1 - q_{sen}$ decreases (indicating increase of multifractality). Multiscaling also increases, as shown especially by the rise in $B$ and, to a lesser extent, the rise in $W$. In S&P 500, we observe a negative value of $B$ and $W$ during the dropping period after 2008, indicating an anomalous 'reversal' of scaling. However, $q_{rel}$ drops for DAX and slightly for NIKKEI, showing the same behavior with endogenous bubbles, with the exception of S&P 500 for which it rises, showing that in S&P 500 'herding' behavior was pronounced during the 2008 crash relative to the period before the crash. The temporal evolution of the $q$-triplet was combined in two single metric quantities: $Q$ and $Q^{inv}$. A large $Q^{inv}$ indicates a big total departure from stochasticity, increased non-extensive statistics as well as pronounced multifractality/multiscaling. A small value of $Q^{inv}$ indicates a behavior close to stochasticity. A large value of $Q$, on the other hand, can be caused by either very increased stochastic behavior (with specially pronounced uniscaling) or very large values of $q_{stat}$ and $q_{rel}$ only. Usually a very small $Q^{inv}$ combined with a large $Q$ was observed in highly-efficient market periods (stochastic behavior) and large $Q^{inv}$ combined with smaller or equal $Q$ was observed in the rising periods of endogenous stock market bubbles or after exogenous market crashes. These metrics appear to be useful as single measure of non-extensiveness and may convey useful information about the market dynamics.

## CONCLUSIONS

In this study we have presented results for time-dependent Tsallis non-extensive statistics by breaking stock market timeseries comprising of daily close prices of four major global indices into several distinct, non-overlapping time periods and estimating the Tsallis $q$-triplet for each one of them. We also presented results for the time-dependent (multi)scaling characteristics of these timeseries by using the GHE method and estimating the width $W_{q,q'}$ and depth $B_{q,q'}$ of the GHE spectrum for a different range of parameter $q$ values ($q = 0.1$, $q' = 4$ for $W$ and $q = 0.1$, $q' = 1$ for $B$) for each of the time periods. Finally, we combined the values of the $q$-triplet for each time period and introduced two metrics of 'non-extensiveness' of the statistics, the $Q$ and $Q^{inv}$ metrics.

We conclude that in financial markets, extensive statistics clearly changes with time. All members of the $q$-triplet as well as their temporal changes (local trends) are useful in conveying relative information about the market evolution, such as the development of a stock market bubble. Particular long-term and local temporal patterns (trends) were observed in the relative changes of the values of each member of the q-triplet with time in the periods before an endogenous stock market bubble. Also, the correlations among $(1 - q_{sen})^{-1}$ $W'$ and $B'$ distinguish between periods just before and just after a bubble break, as well as between an *endogenous* and *exogenous* stock-market crises for the markets studied. Based on the above observations, the present study provided preliminary evidence that the $q$-triplet together with the $W'$ and $B'$ multifractal measures, are (i) able to distinguish between endogenous and exogenous generated critical periods and (ii) possibly provide signals for a developing bubble as well as the approximate time that it is about to break. Also, it is useful to combine the evolution of the $q$-triplet with a measure of multiscaling coming from the GHE method, namely the width $W$ and the depth $B$ (B-proxy), using different ranges of the parameter $q$ for the generalized Hurst, in order to capture different aspects of the price difference distributions. Results of this study seem to be in accordance with previous studies of the temporal evolution of scaling in financial timeseries using only time-dependent Hurst exponents. The increase of $q_{stat}$ in the rising period before stock market bubbles is in agreement with the fattening of the tails of price change distributions in these eras as reported by other studies. Finally, 'herding' in trading behavior during some period is probably related to the value of $q_{rel}$ in that period.

The combined analysis of financial timeseries by the non-extensive statistics of Tsallis and the GHE approach strengthens the understanding of the underlying processes of the economic systems producing the series. The combination of complexity metrics creates new data and conclusions in the direction of a deeper understanding of financial markets. As future work, it would be instructive to see the temporal evolution of the *q*-triplet of Tsallis in combined with the GHE approach using sliding (overlapping) time windows (as was done often in previous time-dependent GHE studies), as this would present the temporal evolution of non-extensive statistics with much greater resolution. This, together with the study of yet more stock market indices is necessary in order to provide further evidence in order to support the main findings of this study. Moreover, information retrieved from such non-linear analysis methods can be exploited by machine learning models for market classification and prediction, allowing more accurate, meaningful and relevant correlation to underlying market conditions. The authors of this work are now working in this direction.

## AUTHOR CONTRIBUTIONS
I.P.A. Conceptualization; I.P.A., L.P.K., E.G.P. Data curation; I.P.A., L.P.K., E.G.P. Formal analysis; I.P.A., L.P.K., E.G.P. investigation; I.P.A., L.P.K., E.G.P. Methodology; I.P.A., L.P.K., E.G.P. Software; I.P.A. validation; I.P.A., L.P.K., E.G.P. Writing – original draft; I.P.A., L.P.K., E.G.P. Writing – review & editing.

## DECLARATION OF INTERESTS
The authors declare no competing interests.